\documentclass[aps,pra,superscriptaddress,twocolumn,twoside,a4paper,showpacs,floatfix]{revtex4}

\usepackage{dcolumn}
\usepackage{enumerate}
\usepackage{bbm}
\usepackage{bm}
\usepackage[french, english]{babel}
\usepackage{synttree}

\usepackage{hyperref,graphicx,amsmath,latexsym,revsymb,amssymb,verbatim,color}

\newcommand{\tr}{\textrm{tr}}
\newcommand{\bra}[1]{\langle #1|}
\newcommand{\ket}[1]{|#1\rangle}

\renewcommand\appendix{\par
  \setcounter{section}{0}
  \setcounter{subsection}{0}
  \setcounter{figure}{0}
  \setcounter{table}{0}
  \renewcommand\thesection{Appendix \Alph{section}}
  \renewcommand\thefigure{\Alph{section}\arabic{figure}}
  \renewcommand\thetable{\Alph{section}\arabic{table}}
}

\begin{document}

\title{Reliable experimental quantification of bipartite entanglement\\without reference frames}

\author{Thomas Lawson} \affiliation{LTCI, CNRS -- T\'el\'ecom ParisTech, Paris, France}
\author{Anna Pappa}  \affiliation{LTCI, CNRS -- T\'el\'ecom ParisTech, Paris, France}\affiliation{LIAFA, CNRS -- Universit\'e Paris 7, Paris, France}
\author{Boris Bourdoncle} \affiliation{LTCI, CNRS -- T\'el\'ecom ParisTech, Paris, France}
\author{Iordanis Kerenidis} \affiliation{LIAFA, CNRS -- Universit\'e Paris 7, Paris, France}
\author{Damian Markham}  \affiliation{LTCI, CNRS -- T\'el\'ecom ParisTech, Paris, France}
\author{Eleni Diamanti}  \affiliation{LTCI, CNRS -- T\'el\'ecom ParisTech, Paris, France}

\date{\today}
\begin{abstract}
Simply and reliably detecting and quantifying entanglement outside laboratory conditions will be essential for future quantum information technologies. Here we address this issue by proposing a method for generating expressions which can perform this task between two parties who do not share a common reference frame. These reference-frame-independent expressions require only simple local measurements, which allows us to experimentally test them using an off-the-shelf entangled photon source. We show that the values of these expressions provide bounds on the concurrence of the state and demonstrate experimentally that these bounds are more reliable than values obtained from state tomography since characterizing experimental errors is easier in our setting. Furthermore, we apply this idea to other quantities, such as the Renyi and von Neumann entropies, which are also more reliably calculated directly from the raw data than from a tomographically reconstructed state. This highlights the relevance of our approach for practical quantum information applications that require entanglement.
\end{abstract}

\maketitle
%%%%%%%%%%%%%%%%%%%%%%%%%%%%%%%%%%%%%%%%%%%%%%%%%%%%%%%%%%%%%%%%%%%%%%%%

\section{Introduction}
Central to the field of quantum information is quantum entanglement \cite{Horodeckis}, a resource which promises to revolutionize many information-theoretic tasks. Quantum technologies for performing these tasks are maturing quickly. Before long the ability to generate and quantify entanglement outside the laboratory will be essential. One problem that appears when moving into real-world conditions is that it is often difficult to establish a common reference frame between distant parties that wish to communicate. A natural question then is whether it is possible to detect and quantify entanglement in this setting.
Several works have addressed this question in the recent years, proposing elegant schemes enabling the detection of entanglement in the absence of a common reference frame \cite{Weinfurter2012, Bjork2007, Kothe2007, Aschauer2004, Laskowski2011, Badziag2008, Vincente2011}. In addition, some of these schemes can quantify the entanglement in the system \cite{Bjork2007, Kothe2007}.

In this work, we examine this question both in theory and in practice, and propose a set of reference frame independent quantities allowing for a simple experimental quantification of bipartite entanglement. In particular, we first present a method for generating expressions that are independent of a shared reference frame and involve only standard local measurements. We analyze the properties of four of these expressions, and for the first one, we show how it can provide tight upper and lower bounds for a widely used entanglement measure, the concurrence \cite{Peres91}.
Furthermore, the simplicity of our scheme allows us to experimentally calculate the expressions using an off-the-shelf entangled photon source. Crucially, we demonstrate that our experimental results provide bounds on the concurrence that are more reliable than values obtained using state tomography.
Indeed, state tomography introduces errors that are difficult to quantify. In contrast, in our expressions, being simple functions of raw data, experimental errors can be reliably calculated.
We also apply this technique to other quantities that are normally calculated from a tomographically reconstructed state, such as the Renyi and von Neumann entropies.
This work therefore provides a reliable means to measure entanglement -- and other quantities --  in realistic conditions, where a shared reference frame may be difficult to establish.

\section{A reference frame independent expression}

The expectation values of measurements on a quantum state can be combined to form expressions that are invariant under local rotation. We will call such expressions reference frame independent (\emph{rfi}), a term that has been first introduced in the context of quantum key distribution \cite{Laing2010}.

We start our analysis of \emph{rfi} expressions by considering a bipartite qubit state, $\rho$, which may be decomposed in the Pauli basis as follows:
\begin{equation}\label{eq: bipartite rho pauli decomposition}
 \rho = \frac{1}{4} \sum_{i,j =0}^3 \langle\sigma_{i} \sigma_{j}\rangle \big(\sigma_{i} \otimes \sigma_{j}\big),
\end{equation}
where $\sigma_0 = I$, the identity operator, $\sigma_1 = \ket{0}\bra{1}+\ket{1}\bra{0}$, $\sigma_2 = i\ket{0}\bra{1}-i\ket{1}\bra{0}$, $\sigma_3 = \ket{0}\bra{0}-\ket{1}\bra{1}$ are the Pauli operators, and by $\langle\sigma_{i} \sigma_{j} \rangle$ we denote the expectation value $\tr(\sigma_{i} \sigma_{j} \rho)$. We can then express the quantity $\tr \rho^2$, \emph{i.e.}, the purity of the state, in this decomposition, as
\begin{align}\label{eq: bipartite purity}
\tr \rho^2
=&\frac{1}{16}\sum_{i,j,i', j'=0}^3  \!\!\!\!
\langle\sigma_{i} \sigma_{j}\rangle \langle\sigma_{i'} \sigma_{j'} \rangle
\tr (\sigma_{i} \otimes\sigma_{i'}) (\sigma_{j} \otimes\sigma_{j'}) \notag\\
=&\frac{1}{4} \sum_{i,j =0}^3 \langle\sigma_{i} \sigma_{j} \rangle^2,
\end{align}
where only non-traceless products, arising for $i=i'$ and $j=j'$, contribute to the summation. Removing all the identity operators from this summation leads to the expression
\begin{equation}\label{eq: Q}
Q_2:= \sum_{i,j =1}^3 \langle\sigma_{i} \sigma_{j}\rangle^2,
\end{equation}
which is known to be a reference frame independent quantity \cite{Aschauer2004, Weinfurter2012, Badziag2008}, meaning that it satisfies
\begin{equation}
Q_2(\rho) = Q_2\big( (R_A \otimes R_B) \rho (R_A \otimes R_B)^{\dagger} \big),
\end{equation}
for any local single qubit rotations, $R_A$ and $R_B$. The quantity $Q_2$ is sensitive to entanglement, taking the value 3 for a maximally entangled state and a value less than or equal to 1 for separable states (the equality holds for a separable pure state). The identity
\begin{align}\label{eq: bipartite full sum}
Q_2 \equiv \sum_{i,j=0}^3 \langle\sigma_{i} \sigma_{j}  \rangle ^2
 - \Big( \sum_{i=0}^3 \langle\sigma_{i} \sigma_{0}  \rangle ^2 + \sum_{j=0}^3 \langle\sigma_{0} \sigma_{j}  \rangle ^2\Big)
+ \langle\sigma_{0} \sigma_{0}  \rangle ^2
\end{align}
allows $Q_2$ to be expressed in terms of full and partial state purities \cite{Aschauer2004},
\begin{equation}\label{eq:Q traces}
Q_2(\rho) = 4 \tr \rho^2-2 \big( \tr \rho_A^2 +  \tr \rho_B^2\big)+ 1,
\end{equation}
where $\rho_{A,B} = \tr_{B,A} (\rho)$. Expressing $Q_2$ as a function of purities shows that it is invariant under local rotations $R_A$ and $R_B$.
It also explains why $Q_2$ detects entanglement, since the entanglement of a bipartite state can be characterized by the full and partial purities \cite{Bovino2005}.
Eq. \eqref{eq:Q traces} can also be written
\begin{equation}
Q_2(\rho) = 4 \tr \rho^2-2 \big(  Q_1(\rho_A) +  Q_1(\rho_B)\big)- 1,
\end{equation}
where $Q_1= \sum_{i=1}^3  \langle\sigma_{i} \rangle ^2$.

References \cite{Aschauer2004, Weinfurter2012, Badziag2008} showed that $Q_2$ can be used to witness entanglement, and also to establish the experimental Schmidt decomposition. We will show in the following that it can also be used to derive bounds on the concurrence, a measure of entanglement \cite{Peres91}.

Before we do this, we present a general method for generating \emph{rfi} expressions.

\section{Method for generating reference frame independent expressions}

Instead of considering the purity, $\tr \rho^2$, let us apply the above argument to the quantity $\tr \rho^n$.
For each power of $n$ ($n \geq 3$), a new $\emph{rfi}$ entanglement sensitive quantity is found by removing identity terms and repetitions of existing expressions from the Pauli decomposition.

For instance, expressing $\tr \rho^3$ in the Pauli decomposition,
\begin{align}\label{eq: tr rho 3}
 \frac{1}{64} \!\!\!\!\! \sum_{i,j,k,l,m,n=0}^3 \!\!\!\!\!  \langle \sigma_i \sigma_j \rangle \langle \sigma_k \sigma_l \rangle \langle \sigma_m \sigma_n \rangle \tr \Big(  (\sigma_i \sigma_k \sigma_m ) \otimes (\sigma_j \sigma_l \sigma_n ) \Big),
\end{align}
and following a straightforward but long calculation (see \ref{Appendix: bipartite rho^3} for details), leads to the expression
\begin{align}
Q_3 := \mbox{ }  & \langle \sigma_1 \sigma_3 \rangle \langle \sigma_2 \sigma_2 \rangle \langle \sigma_3 \sigma_1 \rangle -
\langle \sigma_1 \sigma_2 \rangle \langle \sigma_2 \sigma_3 \rangle \langle \sigma_3 \sigma_1 \rangle \notag\\
- &\langle \sigma_1 \sigma_3 \rangle \langle \sigma_2 \sigma_1 \rangle \langle \sigma_3 \sigma_2 \rangle +
\langle \sigma_1 \sigma_1 \rangle\langle \sigma_2 \sigma_3 \rangle \langle \sigma_3 \sigma_2 \rangle \notag\\
+ &\langle \sigma_1 \sigma_2 \rangle \langle \sigma_2 \sigma_1 \rangle \langle \sigma_3 \sigma_3 \rangle -
\langle \sigma_1 \sigma_1 \rangle \langle \sigma_2 \sigma_2 \rangle \langle \sigma_3 \sigma_3 \rangle,
\end{align}
which is reference frame independent.
The quantity $Q_3$ takes the value 1 for a maximally entangled state and 0 for a separable pure state. As with $Q_2$, we can write $Q_3$ in a more useful way
\begin{align}
  6 Q_3(\rho) = &16 \text{ } \tr \rho^3 - 24 \tr \rho^2 + 3 G(\rho) \notag\\
  + &12 \big( \tr \rho_A^2 + \tr \rho_B^2   -  \tr \rho_A^2 \tr \rho_B^2 \big)  -4 ,
\end{align}
in terms of partial purities and a previously known \emph{rfi} quantity, $G = \sum_{i,j=1}^3 \big(\langle \sigma_i \sigma_j \rangle - \langle \sigma_i \sigma_0 \rangle \langle \sigma_0 \sigma_j \rangle \big)^2$ \cite{Bjork2007, Kothe2007}.
This shows that $Q_3$ is, indeed, \emph{rfi}.

Following the same procedure for higher powers of $\rho$ gives more entanglement sensitive expressions.
As the power increases the calculations become more complicated and expressions must be checked for reference frame independence since they are not always functions of known expressions.
From the decomposition of $\tr \rho^4$ we find the expression
\begin{align}
Q_4:= \!\!\!\! \sum_{i\neq k (i,k \neq 0)}\sum_{j\neq l (j,l \neq 0)}  \!\!\!\! &\langle \sigma_i \sigma_j  \rangle^2 \langle \sigma_k \sigma_l  \rangle^2 - \notag\\
 &  \langle \sigma_i \sigma_j  \rangle \langle \sigma_k \sigma_l  \rangle  \langle \sigma_i \sigma_l  \rangle  \langle \sigma_k \sigma_j  \rangle,
\end{align}
which can be written as a function of other \emph{rfi} expressions
\begin{align}
2 Q_4(\rho) = \text{ } & 64 \tr \rho^4 + 12 G(\rho) \notag\\
- & 2Q_2(\rho)\big( Q_1(\rho_{A}) +  Q_1(\rho_{B}) \big) \notag\\
-&18 Q_1(\rho_{A}) Q_1(\rho_{B}) -6 \big( Q_1(\rho_{A}) +  Q_1(\rho_{B}) \big) \notag\\
- & \big( Q_1^2(\rho_{A}) +  Q_1^2(\rho_{B}) \big)
- Q_2^2(\rho) - 18 Q_2(\rho) \notag\\
+ & 4Y(\rho) - 4 \big(Z_1(\rho)+ Z_2(\rho)\big) \notag\\
- &24   Q_3(\rho) -1,
\end{align}
including $Q_1$, $Q_2$, $Q_3$, and $G$, and three new expressions,
\begin{align}
Y:=\!\!\!\!\!\!\!\!  \mathop{\sum_{i \neq j \neq k (> 0),}}_{ l \neq m \neq n (> 0)} \!\!\!\!\!\!\!
(-1)^{(k-i) \!\!\!\! \mod 3 }  (-1)^{(n-l) \!\!\!\! \mod 3 } \times \notag\\
\langle \sigma_0 \sigma_l \rangle \langle \sigma_i \sigma_0 \rangle \langle \sigma_j \sigma_m \rangle \langle \sigma_k \sigma_n \rangle, \label{eq: Y} \\
Z_1:= \sum_{i,j,k=1}^3  \langle \sigma_0 \sigma_j  \rangle \langle \sigma_0 \sigma_k  \rangle  \langle \sigma_i \sigma_j \rangle \langle \sigma_i \sigma_k \rangle,  \label{eq: X}\\
Z_2:= \sum_{i,j,k=1}^3 \langle \sigma_j \sigma_0  \rangle  \langle \sigma_k \sigma_0  \rangle  \langle \sigma_j \sigma_i  \rangle  \langle \sigma_k \sigma_i  \rangle. \label{eq: X-}
\end{align}
The quantity $Q_4$ takes the value 6 for a maximally entangled state and 0 for a separable pure state. Beyond $n=4$ the expressions become large and writing them down is cumbersome. Nonetheless, more entanglement sensitive expressions exist: $\tr \rho^5$ defines (at least one) new expression, $Q_5$, written explicitly in \ref{Appendix: bipartite rho^5}.
The quantity $Q_5$ takes the value 3 for maximally entangled states and 0 for separable pure states.
It is a simple but long exercise in algebra -- not included here -- to show that $Q_4$ (and $Y$, $Z_1$ and $Z_2$) and $Q_5$ are indeed \emph{rfi} .

\section{Quantifying bipartite entanglement}

We now show that our reference frame independent expressions give bounds on the concurrence, $C$, of a bipartite state \cite{Peres91}.

We start with the lower bound, using the result
\begin{align}\label{eq:concurrence bound}
C^2 \geq 2 \max_{r=A,B} \lbrace \tr \rho^2 - \tr \rho_r^2 \rbrace,
\end{align}
developed in the context of \emph{direct measurements} \cite{Mintert2007}, a way of characterizing entanglement using two-fold copies of the state, $\rho \otimes \rho$ \cite{Ekert2002, Ma2009, Zhang2008}.
By the observation that $2 \max_{r=A,B} \lbrace \tr \rho^2 - \tr \rho_r^2 \rbrace \geq  2\tr \rho^2 - (\tr \rho_A^2+\tr \rho_B^2)$ and Eq. \eqref{eq:Q traces} we find
\begin{align}\label{eq: upper bound on Q}
C^2 \geq  \frac{Q_2-1}{2}.
\end{align}
Note that this bound is only slightly looser that the bound in Eq. \eqref{eq:concurrence bound} since for the (generally, highly entangled) states that interest us the difference $| \tr \rho_A^2 - \tr \rho_B^2|$, is typically small.
Furthermore, although the right hand side of Eq. \eqref{eq:concurrence bound} is \emph{rfi}, the direct measurement methods for the experiments finding it do need aligning, unlike the measurements we use to verify $Q_2$.

It is also possible to derive an upper bound on $C$ based on direct measurement schemes. In this case, the result \cite{Zhang2008}
\begin{align}\label{eq:concurrence lower bounds}
2 \min_{r=A,B} \lbrace 1 - \tr \rho_r^2 \rbrace \geq C^2,
\end{align}
implies the bound $(Q_2 +3 -4  \tr \rho^2)/2 \geq C^2$.
Note that the left hand side of this expression can be calculated from the measurements needed for $Q_2$ since $\tr \rho^2$ can be computed from Pauli measurements, as shown by Eq. \eqref{eq: bipartite purity}.
Unfortunately this bound is loose. Indeed, it cannot be tighter that the bound in Eq. \eqref{eq:concurrence lower bounds}, which is also fairly loose (numerical simulations show that this bound is usually much looser than the one we will propose -- see Section \ref{sec:extensivon}).

We propose instead a bound which is tighter, and justify it with a strong albeit not general argument.
Observing that the lower bound is saturated by a pure state -- and taking inspiration from references \cite{Bjork2007, Kothe2007} --
we assume that our bound is saturated by a state whose mixedness can be varied independently of its entanglement.
Such states, known as \emph{maximally entangled mixed states} (MEMS), were suggested in reference \cite{Wei2003}, 
\begin{align}
 \label{eq: general state Munro}
\rho_{\text{MEMS}}:=\left( \begin{array}{cccc}
x + \frac{\gamma}{2} &0 & 0 & \frac{\gamma}{2}\\
0 &\alpha & 0 & 0\\
0 &0 & \beta & 0\\
 \frac{\gamma}{2} &0 & 0 & y+ \frac{\gamma}{2}
\end{array} \right),
\end{align}
where $x$, $y$, $\alpha$, $\beta$ and $\gamma$ take real, non-negative values such that $\rho_{\text{MEMS}}$ is a valid quantum state.
The concurrence of $\rho_{\text{MEMS}}$ is $C(\rho_{\text{MEMS}}) = \gamma - 2\sqrt{\alpha \beta}$.

Our goal is to minimize $Q_2(\rho_{\text{MEMS}})$ for a given $C$.
Setting $\beta=0$ maximizes the concurrence with no effect on $Q_2(\rho_{\text{MEMS}})$.
By setting $\phi = x+y$ and applying the normalization condition $\phi + \gamma + \alpha= 1$ we find
\begin{align}
Q_2(\rho_{\text{MEMS}}) = 1 + 2 \gamma^2 - 4 \big(1- (\phi + \gamma)\big) (\phi + \gamma).
\end{align}
We then minimize over $\phi$, which leads to the upper bound
\begin{align}\label{eq: lower bound on Q}
Q_2(\rho_{\text{MEMS}})\geq \left\{
\begin{array}{c l}
    2 C^2 & C\leq \frac{1}{2}\\
   1 - 4 C + 6 C^2 & C > \frac{1}{2}
\end{array}\right.
\end{align}

%**************************************************
\begin{figure}[tb]
\centering
\includegraphics[scale=0.43]{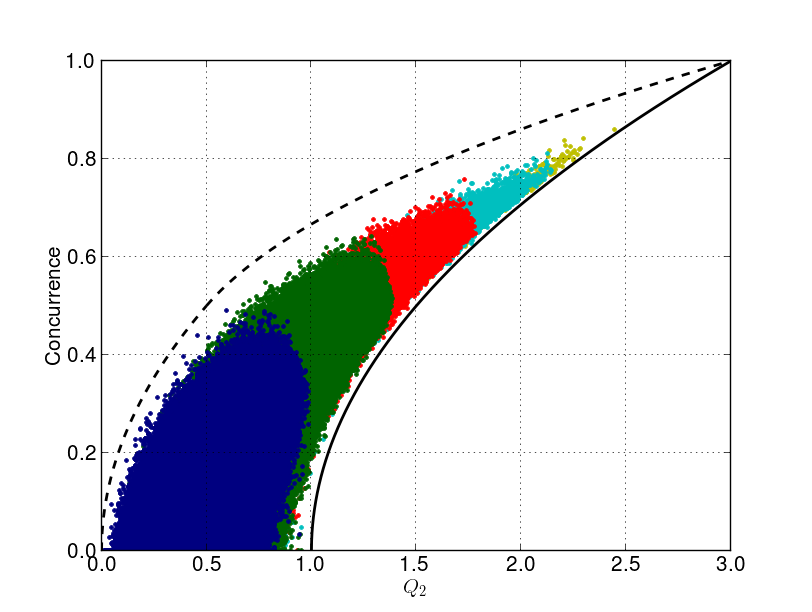}
\caption{(Color online.) Concurrence vs. $Q_2$. 
Several million randomly generated mixed states respect the proposed upper and lower bounds on $C$ (the former plotted in dashed line, since it is not proven in generality). 
Furthermore, categorizing these states by purity lets us further discriminate entanglement. States of purity $\leq 0.5$ are shown in blue, $0.5-0.6$ in green, $0.6-0.7$ in red, $0.7-0.8$ in cyan, and $0.8-0.9$ in yellow (in black and white these categories appear as increasingly pale shades of grey). The boundaries suggest that purity can be used to tighten the upper bound on concurrence. }
\label{fig:numerics of Q vs C}
\end{figure}
%**************************************************

%**************************************************
\begin{figure}[tb]
\centering
\includegraphics[scale=0.43]{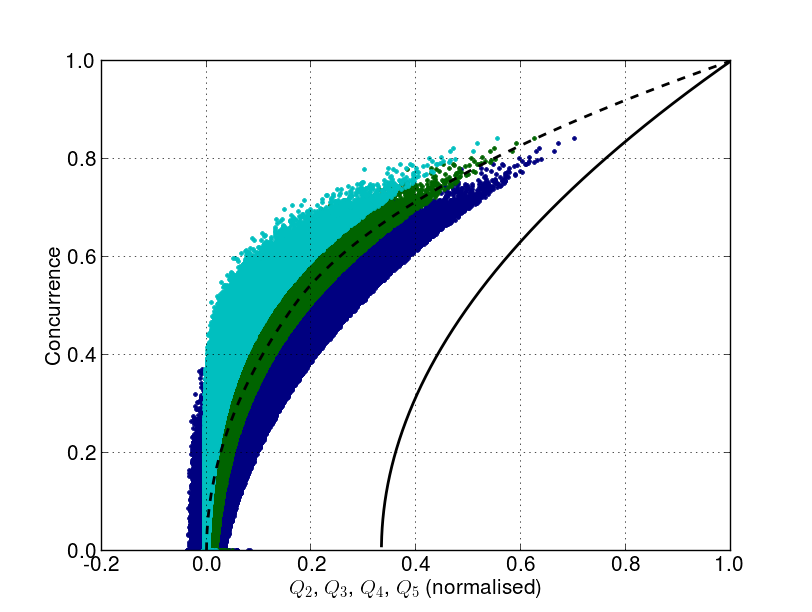}
\caption{(Color online.) Concurrence vs. $Q_3$ (blue), $Q_4$ (green) and $Q_5$ (cyan) (in black and white $Q_3$, $Q_4$, and $Q_5$ are shown in increasingly pale shades of grey) for 5 million randomly generated bipartite mixed states. 
Also shown are the bounds of $Q_2$ (black). All the expressions have been normalized to have maximum value 1.
The spreads of values of $Q_3$, $Q_4$ and $Q_5$ get increasingly narrower, suggesting that $Q_3$, $Q_4$ and $Q_5$  may give tighter bounds on the concurrence than $Q_2$, especially for highly entangled states.}
\label{fig:Q2 S2 P2 R2 against concurrence}
\end{figure}
%**************************************************

The upper and lower bounds for $C$, Eqs. \eqref{eq: lower bound on Q} and \eqref{eq: upper bound on Q}, respectively, are shown in Fig. \ref{fig:numerics of Q vs C}. They are respected by randomly generated mixed states.
A value of $Q_2 >1$ here implies non-separability. A useful property of $Q_2$ is that it lets us lower bound $C$ \emph{efficiently} \cite{Weinfurter2012}, \emph{i.e.} using fewer measurements than needed for state tomography, since any subset, $\mathbb{S}$, of the tomographic measurements lower bounds $Q_2$, $\sum_{i,j \in \mathbb{S}} \langle\sigma_{i} \sigma_{j} \rangle^2 \leq Q_2$.

Reference \cite{Kothe2007} presented bounds on the concurrence in terms of the \emph{rfi} quantity $G$, although these bounds were not proven in generality (and the class of states they use is less general than $\rho_{\text{MEMS}}$). In contrast, our lower bound for $C$ is proven in the general case.
Following the line of enquiry of reference \cite{Kothe2007} we note that the state purity, which is also a \emph{rfi} quantity, may help to quantify entanglement. The mixed states that are used as numerical evidence in Fig. \ref{fig:numerics of Q vs C} are categorized by purity. The boundaries of each category suggest that knowing the purity may help bound the concurrence more tightly. For instance, states of purity $\leq 0.5$ may never be able to cross the $Q_2=1$ boundary, even if they are entangled. This implies that $Q_2$ is not optimal and justifies the search for better \emph{rfi} quantities sensitive to entanglement.

In Fig. \ref{fig:Q2 S2 P2 R2 against concurrence}, we show the values of the \emph{rfi} expressions $Q_3$, $Q_4$ and $Q_5$ for a large number of randomly generated mixed states.
The results suggest that these expressions can quantify bipartite entanglement; indeed, the narrow spreads of their values suggest that they may give tighter bounds on concurrence than $Q_2$, especially for highly entangled states. 
Although we do not prove this, we expect that the \emph{rfi} expressions arising from $\rho^n$, where $n>5$, may give increasingly tight bounds on the concurrence since the quantities $\tr \rho^n$ contain increasingly large amounts of information about the state, letting one calculate -- for instance -- the Renyi entropies (see Section \ref{sec:extensivon}), which are measures of entanglement.

\section{Experimental demonstration}
Our quantities can be calculated using the simple experimental procedure of measuring the Pauli operators on a bipartite entangled state. This allows us to demonstrate the main ideas of this work experimentally using an off-the-shelf entangled photon source \cite{QuTools}. This source generates polarization entangled photon pairs in the state $\ket{\phi^-}=\frac{1}{\sqrt{2}}(\ket{HH}-\ket{VV})$, at a wavelength of 810 nm. Measurements are performed using a rotating quarter wave plate and a polarizer placed at the path of each photon before a silicon avalanche photodiode. The fidelity of the generated state with respect to the maximally entangled state $\ket{\phi^-}$ was $91\%$. 
(Since the model we assume is collaborative, rather than adversarial, as in quantum key distribution, losses play no role -- we consider only photon coincidences.) 

\begin{figure}[tb]
\centering
\includegraphics[scale=0.6]{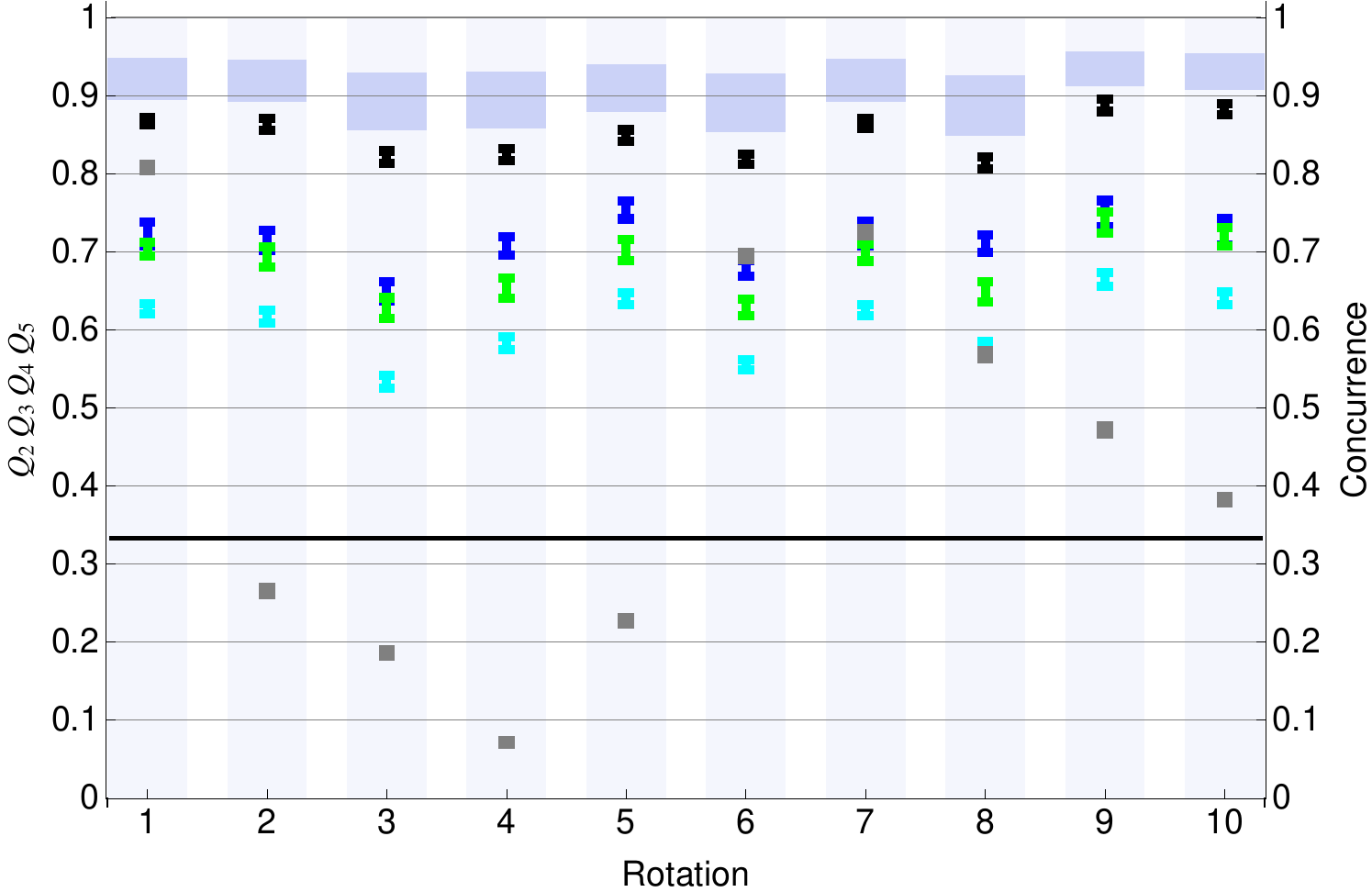}
\caption{The experimental values of $Q_2$, $Q_3$, $Q_4$, $Q_5$ (black, blue, green, cyan -- color online) for ten different rotations of the state. Rotation 1 corresponds to the unrotated state. All quantites have been normalized to a maximal value of 1. The horizontal black line indicates the bound for separable states for $Q_2$, which is equal to $1/3$ because of the normalization. In grey, we show the efficient lower bounds on $Q_2$ found using three measurements chosen according to the method of \cite{Weinfurter2012}. The dark blue regions show possible values of the concurrence, $C$, which is plotted on the right hand axis.}
\label{fig: result bipartite q-t}
\end{figure}

We calculated the values of $Q_2$, $Q_3$, $Q_4$ and $Q_5$ for ten states: the initial unrotated state and nine rotated states, where in each case a randomly chosen rotation was applied to one qubit. Note here that a random rotation on one qubit of a maximally entangled state has the same effect as random rotations on both qubits. The rotations are chosen using the Haar measure, which ensures that they are evenly distributed, and they are applied by adjusting the quarter wave plate and the polarizer in the path of the rotated photon. The results are shown in Fig. \ref{fig: result bipartite q-t}. In all cases the values violate the bounds for separable states, however they do not reach their maximum values because the state is not maximally entangled.

We have also demonstrated that $Q_2$ can be bounded efficiently using the procedure of reference \cite{Weinfurter2012}, which shows how to pick measurements based on previous results in order to prove that an unknown state is entangled using the fewest measurements possible.
The results are shown in Fig. \ref{fig: result bipartite q-t} for the case of three measurements.
Three measurements are sufficient to violate the separability bound of $Q_2$ for most of the rotated states that we have examined. But the method of reference \cite{Weinfurter2012} only guarantees a violation for maximally entangled states, and thus sometimes fails for imperfect states as we observe in Fig. \ref{fig: result bipartite q-t}.

%**************************************************
%**************************************************

The quantities $Q_3$, $Q_4$ and $Q_5$ cannot be calculated efficiently because they contain negative terms. However, all these \emph{rfi} quantities have the crucial advantage of being simple functions of raw data, which means that experimental errors can be easily traced through the calculation and expressed as error bars on the values of the quantities. In contrast, state tomography, which involves matching data to the nearest physical quantum state (statistical variations in the data mean states with negative eigenvalues are often found, for instance), introduces errors which are hard to characterize \cite{Blume-Kohout2012, Christandl2012, Zhu2014, Shang2013}.

We show this experimentally: for each rotation we calculated the upper and lower bounds on the state concurrence, given by Eqs. \eqref{eq: lower bound on Q} and \eqref{eq: upper bound on Q}, respectively, from the measured values of $Q_2$.
The results are shown in Fig. \ref{fig: result bipartite q-t}.
The concurrence of the unrotated state (rotation 1 in Fig. \ref{fig: result bipartite q-t}) is bounded as $0.895 \pm 0.004 \leq C(\rho) \leq 0.948 \pm 0.002$. Note here that the error bars are one standard deviation, found using the standard error propagation formula. This links the values of the expression to the experimental data, which we assume to obey Poissonian statistics.
State tomography using the same data \cite{Tomog} gives a reconstructed state with concurrence $C(\rho_{\text{tomog}})= 0.85$, which suggests that the tomographically reconstructed state is quite different to the real one. Indeed, if we had performed our experiment on $\rho_{\text{tomog}}$, we would have found $Q_2(\rho_{\text{tomog}}) = 2.44$, a significant deviation from the actual (unnormalized) value $Q_2(\rho) = 2.60 \pm 0.01$.
The mismatch between real and reconstructed states has already been reported \cite{Schwemmer2013}, and has led to the development of better tomographic
methods \cite{Blume-Kohout2012, Christandl2012, Zhu2014, Shang2013}, albeit ones which are often difficult to perform in practice.
Expressions such as $Q_2$, $Q_3$, $Q_4$ and $Q_5$, which are simple functions of raw data, provide reliable ways of characterizing entanglement.

\section{Extension to further quantities}\label{sec:extensivon}

This idea can be applied to more complex quantities, which would normally be computed from a tomographically reconstructed state, as was done in reference \cite{Flammia2011}. All functions of a density matrix can be expanded in terms of expectations of the Pauli operators. Expressing them in this form allows us to trace through the experimental errors directly as above.

As an example, let us consider the purity, which can be calculated directly from the tomographic data: the full state purity is given in Eq. \eqref{eq: bipartite purity}, while the partial purity is $\tr \rho_A^2 = 1/2 \sum_{i = 0}^3 \langle \sigma_i \sigma_0 \rangle$. This can be used to implement a simple, but powerful entanglement test. As shown in reference \cite{Bovino2005}, all bipartite separable states obey
\begin{align}\label{eq: bovino measure}
\tr \rho_A^2 \geq \tr \rho^2, \; \; \;\; \;\;\; & \tr \rho_B^2 \geq \tr \rho^2.
\end{align}
Entanglement can thus be proven from the raw data by contradicting this. Our (unrotated) state gives the purities $ \tr \rho_A^2 = 0.5081 \pm 0.0001$,
$ \tr \rho_B^2 = 0.5044 \pm 0.0001$, and $\tr \rho^2 = 0.9066 \pm 0.0008$, comfortably violating the separability criteria.

Similarly, the full and partial state purities can be used to compute the bounds on concurrence proposed in references \cite{Zhang2008, Mintert2007},
\begin{align} \label{eq: con bounds}
2 \max_{r=A,B} \lbrace \tr \rho^2 - \tr \rho_r^2 \rbrace \leq C^2 \leq 2 \min_{r=A,B} \lbrace 1 - \tr \rho_r^2 \rbrace.
\end{align}
References \cite{Ekert2002, Bovino2005} showed that, in principle, the entanglement measures in the Eqs. \eqref{eq: bovino measure} and \eqref{eq: con bounds} can be obtained using just one observable. However, in practice these methods need two-fold copies of the state, $\rho \otimes \rho$, and, in the scheme of \cite{Ekert2002}, a multi-qubit operator controlled by an ancilla qubit;
it is much simpler to generate entangled states one at a time and measure single-qubit observables, which are the only requirements of our method. For our state, we find the bounds $0.8968 \pm 0.0009 \leq C(\rho) \leq 0.9918 \pm 0.0002$. Note here that, while the lower bound agrees with that given by $Q_2$, the upper bound given by Eq. \eqref{eq: con bounds} is much looser. The tomographically reconstructed state gives the bounds $0.85 \leq C(\rho_{\text{tomog}}) \leq 0.99$.

Knowing the purity also lets us bound the Renyi entropies,
\begin{align}
S_{\alpha} := \frac{1}{1-\alpha} \ln \tr \rho^{\alpha}.
\end{align}
Since $S_1 \geq S_2 \geq \ldots \geq S_{\infty}$, knowing $S_2 = - \ln \tr \rho ^{2}$ is enough to upper bound all but the first Renyi entropy, the Von Neumann entropy, $S_1 = - \tr (\rho \ln \rho)$. That is we have
\begin{align}
- \ln \tr \rho^2 = S_2 \geq S_3 \geq  \ldots \geq S_{\infty}.
\end{align}
This can efficiently calculated: one may find a useful bound without performing the full set of tomographic measurements,
\begin{align}
-\ln \Big( \frac{1}{4} \sum_{i,j \in \mathbb{S}} \langle\sigma_{i} \sigma_{j} \rangle^2 \Big) \geq S_2,  
\end{align}
where $\mathbb{S}$ is a subset of the tomographic measurements. (Thus $Q_2$ also bounds $S_2$,  $- \ln Q_2 \geq S_2$.) Our state gives $S_2(\rho) = 0.0981 \pm 0.0009$.
To demonstrate the power of bounding $S_2$ efficiently we consider the largest four expectation values, giving the value $ 0.1188 \pm 0.0009  \geq S_2$, an improvement on the tomographic value, $S_2(\rho_{\text{tomog}})=0.14$, which needs nine measurements.

Of course, the hierarchy of Renyi entropies also implies a lower bound on the von Neumann entropy \cite{Hastings2010}, $S_1 \geq S_2$, although not a very tight one. We now propose a method for finding tighter lower bounds.
We have shown how to calculate $\tr \rho^n$ for $n\geq 2$ from the tomographic data (for instance, Eq. \eqref{eq: tr rho 3} gives $\tr \rho^3$).
Using this we can bound the Von Neumann entropy, which we write as an expectation value, $S_1 = - \langle \ln \rho \rangle$.
We express the natural logarithm as a Mercator expansion,
\begin{align}\label{eq: mercator exp}
S_1 & = \langle 1 - \rho  \rangle + \frac{1}{2} \langle (1 - \rho)^2  \rangle + \frac{1}{3} \langle (1 - \rho)^3  \rangle + \ldots \notag\\
& = \tr \big( \rho ( 1 - \rho ) \big)  + \frac{1}{2} \tr \big( \rho (1 - \rho)^2 \big)   + \frac{1}{3} \tr \big( \rho (1 - \rho)^3  \big) + \ldots.
\end{align}
An abbreviated expansion always lower bounds $S_1$ since each term, $\tr \big( \rho (1 - \rho)^n  \big)/n$, is non-negative.
For instance, given $\tr \rho$, $\tr \rho^2$, $\tr \rho^3$ and $\tr \rho^4$, we can lower bound $S_1$,
\begin{align}
S_1 \geq S_{a} & =  \tr \big( \rho ( 1 - \rho ) \big)  + \frac{1}{2} \tr \big( \rho (1 - \rho)^2 \big)   + \frac{1}{3} \tr \big( \rho (1 - \rho)^3  \big)  \notag\\
& = \frac{11}{6} \tr \rho - 3 \tr \rho^2 + \frac{3}{2} \tr \rho^3 - \frac{1}{3} \tr \rho^4.
 \end{align}
Note that $\tr \rho^n$ is calculated from the same data for any value of $n$. The only limitation in the number of terms in the Mercator expansion is the length of time it takes to compute the expressions. In practice the computation bottleneck arises when calculating error bars.
Using this method our (unrotated) state gives $S_a(\rho) = 0.99 \pm 0.001$, while the Von Neumann entropy calculated from the reconstructed state is $S_1(\rho_{\text{tomog}})= 0.28$.

As shown in reference \cite{Horodecki2002}, given $\tr \nu$, $\tr \nu^2$, $\tr \nu^3$ and $\tr \nu^4$, it is even possible to calculate the eigenvalues of the matrix $\nu$, providing a way to measure entanglement detecting quantities such as concurrence and positivity of the partial transpose (PPT) \cite{Horodeckis} directly.

\section{Discussion}

We have addressed the question of the detection and quantification of bipartite entanglement in the practical setting where a shared reference frame between the two parties is absent by presenting a method for generating expressions that are invariant under reference frame rotation. We have analyzed four expressions and have shown analytically, for the quantity $Q_2$, that it can provide tight bounds to the concurrence of a state, while a numerical analysis suggests that the quantities $Q_3$, $Q_4$ and $Q_5$ may provide even tighter bounds. The quantity $Q_2$ can also be lower bounded using fewer measurements than required for state tomography \cite{Weinfurter2012}. Our expressions have the important advantage with respect to state tomography that they are calculated from raw experimental data, which means that errors can be easily characterized and be represented as error bars on the values of the expressions. Using an off-the-shelf source, we have shown experimentally bounds on concurrence given by these
expressions are more reliable than the values calculated using state tomography.
In this sense, our expressions provide efficient and reliable tests of entanglement for rotated quantum states.

Furthermore, we have applied the same idea to other quantities, which are normally calculated from a tomographically reconstructed state.
Once again, the values of expressions that are simple functions of raw data, for which an analysis of experimental errors is straightforward, are more reliable than those found from a reconstructed state. Our experiment therefore shows that complicated, non-linear quantities can be directly (and sometimes efficiently) estimated from the results of a simple experimental setup. Calculated directly, these quantities are more reliable -- and have errors that are more easily characterized -- than those found from state tomography.

We pose two questions which we leave unanswered. First, it would be interesting to perform this experiment using mixed states -- to see the effect of purity on the reliability of our expressions relative to those derived from state tomography. Second, it would be interesting to derive analytic bounds for the higher order expressions and to find more \emph{rfi} expressions that can be efficiently bounded. Finding such expressions in a multiparty setting is also an important question with implications in the practical demonstration of advanced quantum information protocols.

\section*{Acknowledgments}

We thank QuTools for technical assistance. We acknowledge financial support from the City of Paris through the project CiQWii. I.K. was supported from the ERC project QCC. T.L. and A.P. acknowledge support from Digiteo.\\

%%%%%%%%%%%%%%%%%%%%%%%%%%%%%%%%%%%%%%%%%%%%%%%%%%%%%%%%%%%%%%%%%%%%%%%%

\bibliography{Rfi_multipartite_witness_paper}

%%%%%%%%%%%%%%%%%%%%%%%%%%%%%%%%%%%%%%%%%%%%%%%%%%%%%%%%%%%%%%%%%%%%%%%%
\appendix
\section{Deriving $Q_3$ from $\tr \rho^3$}\label{Appendix: bipartite rho^3}
In this section we apply our method for generating reference frame independent expressions to $\tr \rho^3$.
Consider the decomposition
\begin{align}\label{eq: rho ^3}
 & \tr \rho^3 =\notag\\
 & \frac{1}{64}  \sum_{i,j,k,l,m,n=0}^3 \langle \sigma_i \sigma_j \rangle \langle \sigma_k \sigma_l \rangle \langle \sigma_m \sigma_n \rangle \tr \Big(  (\sigma_i \sigma_k \sigma_m ) \otimes (\sigma_j \sigma_l \sigma_n ) \Big).
\end{align}
Non-traceless operations are defined by $\sigma_i \sigma_k \sigma_m = \sigma_0$. This occurs for:\\

$i)$ $i=k=m=0$;\\

$ii)$ $i=0$, $k=m\neq0$ (and permutations: $i=m\neq0$, $k=0$, etc.); \\

$iii)$ $i \neq k \neq m$ ($i,k,m \neq 0$).\\

These categories account for 256 terms in Eq. \eqref{eq: rho ^3}. Each of these must be examined --
by analyzing all permutations of cases $i)$, $ii)$ and $iii)$ for each party -- to find a \emph{rfi} quantity. \\

\noindent\textbf{Case $iii)$--$i)$ and $iii)$--$ii)$}.\\
If one particle is measured using operators obeying $iii)$, and the other according to either $i)$ or $ii)$, represented in shorthand $iii)$--$i)$ and $iii)$--$ii)$, respectively, then the expectation values combine trivially.
Each element of $\tr \rho^3$ is real (being composed of the product of quantum expectation values). Since measurements according to $iii)$ give imaginary coefficients, but $i)$ and $ii)$ do not, it follows that all combinations such as $iii)$--$i)$ and $iii)$--$ii)$ must be identically zero.\\

We consider now what happens when one party measures according to $i)$. The case $i)$--$iii)$ has just been dealt with, since there is a symmetry between the two parties. This leaves two possibilities.\\

\noindent\textbf{Case $i)$--$i)$}.\\
This defines the term $\langle \sigma_0 \sigma_0 \rangle \langle \sigma_0 \sigma_0 \rangle \langle \sigma_0 \sigma_0 \rangle$, which is 1 for any normalized state.\\

\noindent\textbf{Case $i)$--$ii)$ and $ii)$--$i)$}.\\
This gives three repetitions of the terms
\begin{align}
\sum_{i=1}^3  \langle \sigma_i \sigma_0 \rangle ^2,
\end{align}
and
\begin{align}
\sum_{j=1}^3  \langle \sigma_0 \sigma_j \rangle ^2.
\end{align}
This is equal to $3 (Q_1(\rho_A)+Q_1(\rho_B))$, where $Q_1= \sum_{i=1}^3  \langle\sigma_{i} \rangle ^2$.\\

\noindent\textbf{Case $ii)$--$ii)$}.\\
When $i=j=0$ we find the terms
\begin{align}
&\sum_{j,k=1}^3\langle \sigma_0 \sigma_0 \rangle \langle \sigma_j \sigma_k \rangle \langle \sigma_j \sigma_k \rangle = \sum_{j,k=1}^3  \langle \sigma_j \sigma_k \rangle ^2.
\end{align}
Accounting for repetitions, this gives $3  Q_2(\rho)$. The remaining terms are (six) permutations of
\begin{align}
\sum_{i,j=1}^3  \langle \sigma_i \sigma_0 \rangle \langle \sigma_0 \sigma_j \rangle \langle \sigma_i \sigma_j \rangle.
\end{align}
This can be represented in terms of $G$, an existing \emph{rfi} entanglement measure \cite{Bjork2007, Kothe2007},
\begin{align}
 G =& \sum_{i,j=1}^3  \big( \langle \sigma_i \sigma_j  \rangle -  \langle \sigma_i \sigma_0 \rangle \langle \sigma_0 \sigma_j \rangle  \big)^2\notag\\
 =& \mbox{ }Q_2(\rho)  +  Q_1(\rho_{A}) Q_1(\rho_{B})  - 2 \sum_{i,j=1}^3 \langle \sigma_i \sigma_0 \rangle \langle \sigma_0 \sigma_j \rangle  \langle \sigma_i \sigma_j  \rangle.
\end{align}
The contribution from $ii)$--$ii)$ is thus
\begin{align}
 6 Q_2(\rho)  + 3 Q_1(\rho_{A}) Q_1(\rho_{B}) -3 G.
\end{align}

\vspace{0.1in}
\noindent\textbf{Case $iii)$--$iii)$}.\\
The last contribution -- the one which gives $Q_3$ -- is
\begin{align}\label{eq: rho 3 36 terms expression}
- \sum_{iii) \text{--} iii) }
(-1)^{(m-i) \!\!\!\! \mod 3 }  (-1)^{(n-j) \!\!\!\! \mod 3 }
\langle \sigma_i \sigma_j \rangle \langle \sigma_k \sigma_l \rangle \langle \sigma_m \sigma_n \rangle,
\end{align}
where the summation is over operators defined by $iii)$.

Equation \eqref{eq: rho 3 36 terms expression} is a \emph{rfi} quantity. However, it contains some redundancies; removing repetitions, we find the expression
\begin{align}
Q_3 := &\mbox{ }\langle \sigma_1 \sigma_3 \rangle \langle \sigma_2 \sigma_2 \rangle \langle \sigma_3 \sigma_1 \rangle -
\langle \sigma_1 \sigma_2 \rangle \langle \sigma_2 \sigma_3 \rangle \langle \sigma_3 \sigma_1 \rangle -\notag\\
&\langle \sigma_1 \sigma_3 \rangle \langle \sigma_2 \sigma_1 \rangle \langle \sigma_3 \sigma_2 \rangle +
\langle \sigma_1 \sigma_1 \rangle \langle \sigma_2 \sigma_3 \rangle \langle \sigma_3 \sigma_2 \rangle +\notag\\
&\langle \sigma_1 \sigma_2 \rangle \langle \sigma_2 \sigma_1 \rangle \langle \sigma_3 \sigma_3 \rangle -
\langle \sigma_1 \sigma_1 \rangle \langle \sigma_2 \sigma_2 \rangle \langle \sigma_3 \sigma_3 \rangle.
\end{align}

Using the results of the previous cases, we can write the expression $Q_3$ as
\begin{align}
  6 Q_3 (\rho) = & \mbox{ }16 \tr \rho^3  - 3 Q_1(\rho_{A})  - 3 Q_1(\rho_{B}) - \notag\\
  & 6 Q_2(\rho)  - 3 Q_1(\rho_{A}) Q_1(\rho_{B}) +3 G -  1,
\end{align}
which can be rewritten as
\begin{align}
  6 Q_3(\rho) = & \mbox{ }16 \tr \rho^3 - 24 \tr \rho^2 + 3 G(\rho) +\notag\\
   &12 \big( \tr \rho_A^2 + \tr \rho_B^2   -  \tr \rho_A^2 \tr \rho_B^2 \big)  -4.
\end{align}

%%%%%%%%%%%%%%%%%%%%%%%%%%%%%%%%%%%%%%%%%%%%%%%%%%%%%%%%%%%%%%%%%%%%%%%%

\section{$Q_5$}\label{Appendix: bipartite rho^5}
Our method for generating \emph{rfi} expressions applied to $\tr \rho^5$ gives the expression
\begin{align}
&Q_5:=\notag\\
&\langle \sigma_ 1 \sigma_ 1 \rangle^2 \langle \sigma_ 1 \sigma_ 3 \rangle \langle \sigma_ 2 \sigma_ 2 \rangle \langle \sigma_ 3 \sigma_ 1 \rangle +
\langle \sigma_ 1 \sigma_ 2 \rangle^2 \langle \sigma_ 1 \sigma_ 3 \rangle \langle \sigma_ 2 \sigma_ 2 \rangle \langle \sigma_ 3 \sigma_ 1 \rangle + \notag\\
&\langle \sigma_ 1 \sigma_ 3 \rangle^3 \langle \sigma_ 2 \sigma_ 2 \rangle \langle \sigma_ 3 \sigma_ 1 \rangle + \langle \sigma_ 1 \sigma_ 3 \rangle\langle \sigma_ 2 \sigma_ 1 \rangle^2 \langle \sigma_ 2 \sigma_ 2 \rangle \langle \sigma_ 3 \sigma_ 1 \rangle + \notag\\
&\langle \sigma_ 1 \sigma_ 3 \rangle \langle \sigma_ 2 \sigma_ 2 \rangle^3 \langle \sigma_ 3 \sigma_ 1 \rangle - \langle \sigma_ 1 \sigma_ 1 \rangle^2 \langle \sigma_ 1 \sigma_ 2 \rangle \langle \sigma_ 2 \sigma_ 3 \rangle \langle \sigma_ 3 \sigma_ 1 \rangle - \notag\\
&\langle \sigma_ 1 \sigma_ 2 \rangle^3 \langle \sigma_ 2 \sigma_ 3 \rangle \langle \sigma_ 3 \sigma_ 1 \rangle - \langle \sigma_ 1 \sigma_ 2 \rangle \langle \sigma_ 1 \sigma_ 3 \rangle^2 \langle \sigma_ 2 \sigma_ 3 \rangle \langle \sigma_ 3 \sigma_ 1 \rangle - \notag\\
&\langle \sigma_ 1 \sigma_ 2 \rangle \langle \sigma_ 2 \sigma_ 1 \rangle^2 \langle \sigma_ 2 \sigma_ 3 \rangle \langle \sigma_ 3 \sigma_ 1 \rangle - \langle \sigma_ 1 \sigma_ 2 \rangle \langle \sigma_ 2 \sigma_ 2 \rangle^2 \langle \sigma_ 2 \sigma_ 3 \rangle \langle \sigma_ 3 \sigma_ 1 \rangle + \notag\\
&\langle \sigma_ 1 \sigma_ 3 \rangle \langle \sigma_ 2 \sigma_ 2 \rangle \langle \sigma_ 2 \sigma_ 3 \rangle^2 \langle \sigma_ 3
\sigma_ 1 \rangle - \langle \sigma_ 1 \sigma_ 2 \rangle \langle \sigma_ 2 \sigma_ 3 \rangle^3 \langle \sigma_ 3 \sigma_ 1 \rangle +\notag\\
& \langle \sigma_ 1 \sigma_ 3 \rangle \langle \sigma_ 2 \sigma_ 2 \rangle \langle \sigma_ 3 \sigma_ 1 \rangle^3 - \langle \sigma_ 1
\sigma_ 2 \rangle \langle \sigma_ 2 \sigma_ 3 \rangle \langle \sigma_ 3 \sigma_ 1 \rangle^3 - \notag\\
&\langle \sigma_ 1 \sigma_ 1 \rangle^2 \langle \sigma_ 1 \sigma_ 3 \rangle \langle \sigma_ 2 \sigma_ 1 \rangle \langle \sigma_ 3 \sigma_ 2 \rangle - \langle \sigma_ 1 \sigma_2 \rangle^2 \langle \sigma_ 1 \sigma_ 3 \rangle \langle \sigma_ 2 \sigma_ 1 \rangle \langle \sigma_ 3 \sigma_ 2 \rangle - \notag\\
& \langle \sigma_ 1 \sigma_3 \rangle^3 \langle \sigma_ 2 \sigma_ 1 \rangle \langle \sigma_ 3 \sigma_ 2 \rangle - \langle \sigma_ 1 \sigma_ 3 \rangle \langle \sigma_ 2 \sigma_1 \rangle^3 \langle \sigma_ 3 \sigma_ 2 \rangle - \notag\\
& \langle \sigma_ 1 \sigma_ 3 \rangle \langle \sigma_ 2 \sigma_ 1 \rangle \langle \sigma_ 2 \sigma_2 \rangle^2 \langle \sigma_ 3 \sigma_ 2 \rangle + \langle \sigma_ 1 \sigma_ 1 \rangle^3 \langle \sigma_ 2 \sigma_ 3 \rangle \langle \sigma_ 3 \sigma_ 2 \rangle + \notag\\
& \langle \sigma_ 1 \sigma_ 1 \rangle \langle \sigma_ 1 \sigma_ 2 \rangle^2 \langle \sigma_ 2 \sigma_ 3 \rangle \langle \sigma_ 3 \sigma_ 2 \rangle +\langle \sigma_ 1 \sigma_ 1 \rangle \langle \sigma_ 1 \sigma_ 3 \rangle^2 \langle \sigma_ 2 \sigma_ 3 \rangle \langle \sigma_ 3 \sigma_ 2 \rangle +\notag\\
& \langle \sigma_ 1 \sigma_ 1 \rangle \langle \sigma_ 2 \sigma_ 1 \rangle^2 \langle \sigma_ 2 \sigma_ 3 \rangle \langle \sigma_ 3 \sigma_ 2 \rangle + \langle \sigma_ 1 \sigma_ 1 \rangle \langle \sigma_ 2 \sigma_ 2 \rangle^2 \langle \sigma_ 2 \sigma_ 3 \rangle \langle \sigma_ 3 \sigma_ 2 \rangle - \notag\\
&\langle \sigma_ 1 \sigma_ 3 \rangle \langle \sigma_ 2 \sigma_ 1 \rangle \langle \sigma_ 2 \sigma_3 \rangle^2 \langle \sigma_ 3 \sigma_ 2 \rangle +\langle \sigma_ 1 \sigma_ 1 \rangle \langle \sigma_ 2 \sigma_ 3 \rangle^3 \langle \sigma_ 3 \sigma_ 2 \rangle - \notag\\
&\langle \sigma_ 1 \sigma_ 3 \rangle \langle \sigma_ 2 \sigma_ 1 \rangle \langle \sigma_ 3 \sigma_ 1 \rangle^2 \langle \sigma_ 3 \sigma_ 2 \rangle + \langle \sigma_ 1 \sigma_ 1 \rangle \langle \sigma_ 2 \sigma_ 3 \rangle \langle \sigma_ 3 \sigma_ 1 \rangle^2 \langle \sigma_ 3 \sigma_ 2 \rangle + \notag\\
& \langle \sigma_ 1 \sigma_ 3 \rangle \langle \sigma_ 2 \sigma_ 2 \rangle \langle \sigma_ 3 \sigma_ 1 \rangle \langle \sigma_ 3 \sigma_2 \rangle^2 - \langle \sigma_ 1 \sigma_ 2 \rangle \langle \sigma_ 2 \sigma_ 3 \rangle \langle \sigma_ 3 \sigma_ 1 \rangle \langle \sigma_ 3 \sigma_ 2 \rangle^2 - \notag\\
&\langle \sigma_ 1 \sigma_ 3 \rangle \langle \sigma_ 2 \sigma_ 1 \rangle \langle \sigma_ 3 \sigma_ 2 \rangle^3 + \langle \sigma_ 1 \sigma_ 1 \rangle \langle \sigma_ 2 \sigma_ 3 \rangle \langle \sigma_ 3 \sigma_ 2 \rangle^3 + \notag\\
&\langle \sigma_ 1 \sigma_ 1 \rangle^2 \langle \sigma_ 1 \sigma_ 2 \rangle\langle \sigma_ 2 \sigma_ 1 \rangle \langle \sigma_ 3 \sigma_ 3 \rangle + \langle \sigma_ 1 \sigma_ 2 \rangle^3 \langle \sigma_ 2 \sigma_ 1 \rangle \langle \sigma_ 3 \sigma_ 3 \rangle +\notag\\
& \langle \sigma_ 1 \sigma_ 2 \rangle \langle \sigma_ 1 \sigma_ 3 \rangle^2 \langle \sigma_ 2 \sigma_ 1 \rangle \langle \sigma_ 3 \sigma_ 3 \rangle + \langle \sigma_ 1 \sigma_ 2 \rangle \langle \sigma_ 2 \sigma_ 1 \rangle^3 \langle \sigma_ 3 \sigma_ 3 \rangle -\notag\\
& \langle \sigma_ 1 \sigma_ 1 \rangle^3 \langle \sigma_ 2 \sigma_ 2 \rangle \langle \sigma_ 3 \sigma_ 3 \rangle - \langle \sigma_ 1 \sigma_ 1 \rangle \langle \sigma_ 1 \sigma_ 2 \rangle^2 \langle \sigma_ 2 \sigma_ 2 \rangle \langle \sigma_ 3 \sigma_ 3 \rangle - \notag\\
& \langle \sigma_ 1 \sigma_ 1 \rangle \langle \sigma_ 1 \sigma_ 3 \rangle^2 \langle \sigma_ 2 \sigma_ 2 \rangle \langle \sigma_ 3 \sigma_ 3 \rangle - \langle \sigma_ 1 \sigma_ 1 \rangle \langle \sigma_ 2 \sigma_ 1 \rangle^2 \langle \sigma_ 2 \sigma_ 2 \rangle \langle \sigma_ 3 \sigma_ 3 \rangle +\notag\\
& \langle \sigma_ 1 \sigma_ 2 \rangle\langle \sigma_ 2 \sigma_ 1 \rangle \langle \sigma_ 2 \sigma_ 2 \rangle^2 \langle \sigma_ 3 \sigma_ 3 \rangle - \langle \sigma_ 1 \sigma_ 1 \rangle \langle \sigma_ 2 \sigma_ 2 \rangle^3 \langle \sigma_ 3 \sigma_ 3 \rangle +\notag\\
& \langle \sigma_ 1 \sigma_ 2 \rangle \langle \sigma_ 2 \sigma_ 1 \rangle \langle \sigma_ 2 \sigma_ 3 \rangle^2 \langle \sigma_ 3 \sigma_ 3 \rangle - \langle \sigma_ 1 \sigma_ 1 \rangle \langle \sigma_ 2 \sigma_ 2 \rangle \langle \sigma_ 2 \sigma_ 3 \rangle^2 \langle \sigma_ 3 \sigma_ 3 \rangle  +\notag\\
& \langle \sigma_ 1 \sigma_ 2 \rangle \langle \sigma_ 2 \sigma_ 1 \rangle \langle \sigma_ 3 \sigma_ 1 \rangle^2 \langle \sigma_ 3 \sigma_ 3 \rangle - \langle \sigma_ 1 \sigma_ 1 \rangle \langle \sigma_ 2 \sigma_ 2 \rangle \langle \sigma_ 3 \sigma_ 1 \rangle^2 \langle \sigma_ 3 \sigma_ 3 \rangle +\notag\\
& \langle \sigma_ 1 \sigma_ 2 \rangle \langle \sigma_ 2 \sigma_ 1 \rangle \langle \sigma_ 3 \sigma_2 \rangle^2 \langle \sigma_ 3 \sigma_ 3 \rangle - \langle \sigma_ 1 \sigma_ 1 \rangle \langle \sigma_ 2 \sigma_ 2 \rangle \langle \sigma_ 3 \sigma_ 2 \rangle^2 \langle \sigma_ 3 \sigma_ 3 \rangle + \notag\\
&\langle \sigma_ 1 \sigma_ 3 \rangle \langle \sigma_ 2 \sigma_ 2 \rangle \langle \sigma_ 3 \sigma_ 1 \rangle \langle \sigma_ 3 \sigma_ 3 \rangle^2 - \langle \sigma_ 1 \sigma_ 2 \rangle \langle \sigma_ 2 \sigma_ 3 \rangle \langle \sigma_ 3 \sigma_ 1 \rangle \langle \sigma_ 3 \sigma_ 3 \rangle^2 - \notag\\
&\langle \sigma_ 1 \sigma_ 3 \rangle \langle \sigma_ 2 \sigma_ 1 \rangle \langle \sigma_ 3 \sigma_ 2 \rangle \langle \sigma_ 3 \sigma_ 3 \rangle^2 + \langle \sigma_ 1 \sigma_ 1 \rangle \langle \sigma_ 2 \sigma_ 3 \rangle \langle \sigma_ 3 \sigma_ 2 \rangle \langle \sigma_ 3 \sigma_ 3 \rangle^2 + \notag\\
&\langle \sigma_ 1 \sigma_ 2 \rangle \langle \sigma_ 2 \sigma_ 1 \rangle \langle \sigma_ 3 \sigma_ 3 \rangle^3 - \langle \sigma_ 1 \sigma_ 1 \rangle \langle \sigma_ 2 \sigma_ 2 \rangle \langle \sigma_ 3 \sigma_ 3 \rangle^3.
\end{align}

\end{document}